\documentclass[prl,twocolumn,floatix,amssymb,showpacs,amsmath,superscriptaddress]{revtex4-1}
\pdfoutput=1
\usepackage{graphicx}
\usepackage{epsfig}
\usepackage{amsmath,amssymb}
\usepackage{ dsfont }
\DeclareMathOperator{\sgn}{sgn}
\begin{document}

\newcommand{\ket}[1]{\ensuremath{\left|{#1}\right\rangle}}
\newcommand{\bra}[1]{\ensuremath{\left\langle{#1}\right|}}
\newcommand{\quadr}[1]{\ensuremath{{\not}{#1}}}
\newcommand{\quadrd}[0]{\ensuremath{{\not}{\partial}}}
\newcommand{\slpar}{\partial\!\!\!/}
\newcommand{\gtrescero}{\gamma_{(3)}^0}
\newcommand{\gtresuno}{\gamma_{(3)}^1}
\newcommand{\gtresi}{\gamma_{(3)}^i}

\title{Digital Quantum Simulation of the Holstein Model in Trapped Ions}

\date{\today}

\author{A. Mezzacapo}
\affiliation{Departamento de Qu\'{\i}mica F\'{\i}sica, Universidad del Pa\'{\i}s Vasco UPV/EHU, Apartado 644, 48080 Bilbao, Spain}
\author{J. Casanova}
\affiliation{Departamento de Qu\'{\i}mica F\'{\i}sica, Universidad del Pa\'{\i}s Vasco UPV/EHU, Apartado 644, 48080 Bilbao, Spain}
\author{L. Lamata}
\affiliation{Departamento de Qu\'{\i}mica F\'{\i}sica, Universidad
del Pa\'{\i}s Vasco UPV/EHU, Apartado 644,
48080 Bilbao, Spain}
\author{E. Solano}
\affiliation{Departamento de Qu\'{\i}mica F\'{\i}sica, Universidad del Pa\'{\i}s Vasco UPV/EHU, Apartado 644, 48080 Bilbao, Spain}
\affiliation{IKERBASQUE, Basque Foundation for Science, Alameda Urquijo 36, 48011 Bilbao, Spain}

\begin{abstract}
We propose the implementation of the Holstein model by means of digital methods in a linear chain of trapped ions.  We show how the simulation fidelity scales with the generation of phononic excitations. We propose a decomposition and a stepwise trapped-ion implementation of the Holstein Hamiltonian. Via numerical simulations,  we study how the protocol is affected by realistic gates. Finally, we show how measurements of the size of the simulated polaron can be performed.
\end{abstract}

\pacs{03.67.Lx, 37.10.Ty, 71.38.Ht}

\maketitle

Quantum simulators~\cite{Feynman82, Lloyd96} are promising tools for the deep comprehension of complex quantum dynamics. In a quantum simulator, the higher control on the simulating system can allow to reproduce and recover nontrivial quantum behaviors. Recently, a significant boost to the field of quantum simulations has been provided by the use of digital approximations in trapped-ion setups~\cite{Lanyon11, Casanova12}, based on stroboscopic decompositions of unitary operators~\cite{Suzuki90,Berry07}. However, the digital simulation of coupled bosonic-fermionic systems, naturally described by unbounded Hamiltonians, has not been considered. 

Strongly correlated quantum many body systems represent a challenge to both computational and analytic methods. Among them, correlated fermionic-bosonic models are of critical relevance.
The importance of correlation between electrons and ion vibrations has been proven for a large number of condensed-matter systems~\cite{Alexandrov07}. Their role in high-temperature superconductors, as fullerides and cuprates, is still debated~\cite{Gunnarsson97,Lanzara01,Bulut96}. In solid state systems, the correlation between the presence of electrons in a lattice and deformations of the latter can result in the formation of polarons: electrons and phonons can no longer be considered as stand-alone particles. Depending on the strength of the electron-phonon couplings, the cloud of lattice displacements surrounding the electron can have different sizes. For strong couplings, the electrons can be trapped, with remarkable changes of global properties~\cite{Romero99}. The Holstein model~\cite{Holstein59} has been proved to naturally describe the strong coupling case. This model has been recently addressed by heavy numerical simulations~\cite{Fehske11} and classical analog simulations for a reduced number of sites~\cite{Longhi11}. Perturbation methods based on the Lang-Firsov approximations~\cite{Lang63}, valid in the strong coupling limit, are known since long times. The dimensionality of the underlying lattice also raises critical features ~\cite{Ku02}. While involving a lot of efforts, the full and complete comprehension of the electron-phonon correlations is still an open problem. From a quantum mechanics point of view, when considering creation of phonons, even with few electron sites, the size of the simulated Hilbert space can dramatically grow. The quantum simulation of such a complex dynamics could represent an important step forward in the description of condensed matter systems.

Trapped-ion systems are among the most controllable quantum systems. They offer remarkable computational power to perform quantum simulations exponentially faster than their classical counterparts~\cite{Jane, Porras, Friedenauer08, Kim10, Weimar10, Lamata07, Gerritsma1, Casanova1, Gerritsma2, Casanova11, Lamata11, CasanovaQFT, Schmidt-Kaler, Britton,WunderlichReview, BlattQSim}.

In this Letter, we propose the implementation of the Holstein Hamiltonian in a chain of trapped ions, using digital-analog approximation methods, in which the fermionic part is digitized and the bosonic part is analog and provided naturally by the phonons. First, we address the problem of simulating unbounded Hamiltonians with digital-analog protocols. Then, we provide a convenient decomposition of the Holstein Hamiltonian, in that each step can be implemented in a trapped-ion setup. We discuss a possible experimental implementation, testing the whole protocol with numerical integrations of the Schr\"odinger equation. We show how critical observables, as electron-phonon correlations, can be retrieved from the trapped ion setup, leading to an estimation of the polaron size.

{\it Decomposition of the Model.-} It is known that the dynamics of a quantum state under the action of a Hamiltonian $H$ can be recovered by using combined fractal-stroboscopic symmetric decompositions~\cite{Suzuki90,Berry07}. In most practical cases, one can assume a fractal depth of one. This will be the case through all the rest of our analysis.
With these techniques, the target Hamiltonian $H$ is decomposed in a set of $m$ terms: $H=\sum_{i=1}^mH_i$. Then, the {\it symmetric} decomposition for the unitary operator encoding the dynamics of Hamiltonian $H$ reads
\begin{equation}
U_r(t)=\left(\prod_{i=1}^me^{-\frac{iH_{i}t}{2r}}\prod_{i=m}^1e^{-\frac{iH_{i}t}{2r}}\right)^r.
\end{equation}
Here $r$ is the degree of approximation in terms of Trotter steps. It has been shown~\cite{Berry07} that, using symmetric Suzuki fractal decompositions, the number
of gates needed to approximate the exact time evolution of the quantum state grows with the norm of the simulated Hamiltonian. Therefore, it is a natural problem to think of a quantum simulation involving particle generation, in particular of bosons, whose number can grow, in principle, indefinitely. However, in the standard approach to these problems, the dynamics of a bosonic Hilbert space can be recovered by truncating at a certain point of the number of possible bosonic excitations. Thus, the number of gates needed to achieve a certain fidelity for the simulated quantum state grows as more bosonic excitations are created. 

\begin{figure}[t] 
\includegraphics[width=1\linewidth]{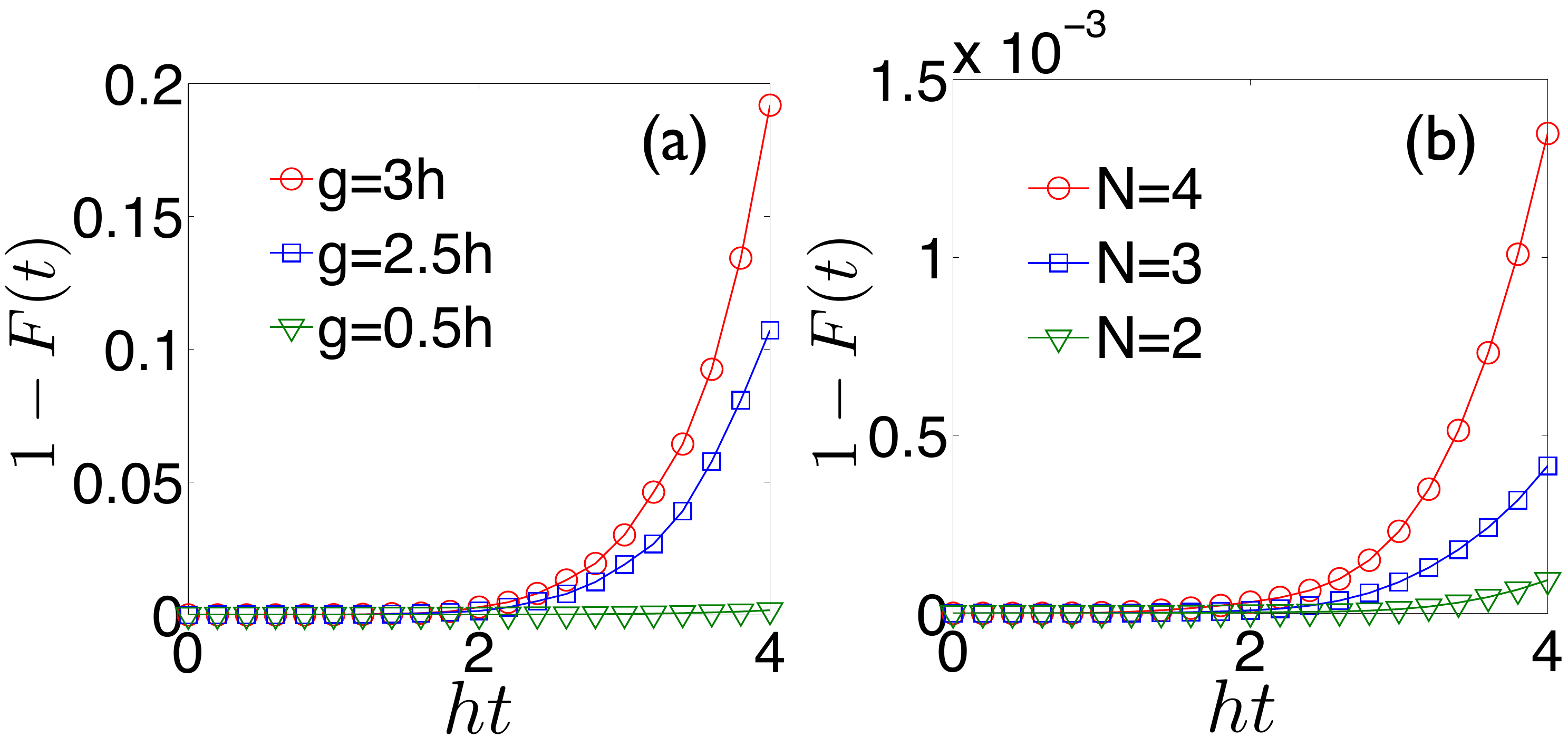}
\caption{(color online). (a)  Behavior of the fidelity loss $1-F(t)=1-|\langle\Psi_E(t)|\Psi_S(t)\rangle|^2$, for a two site configuration, as a function of the electron-phonon coupling strength $g$, for $\omega_0=h/4$. As the coupling $g$ increases, more phonons are created, the Hilbert space describing the dynamics enlarges and the fidelity decreases for a fixed number of approximant gates ($r=10$ here). (b) Dependence of the fidelity loss in the number of sites. Here $g=0.3~h$, $\omega_0=0.5~h$, and ten symmetric steps are considered ($r=10$). The initial state of both plots corresponds to a configuration in which an electron is injected in the site $N/2$ (N even) or $(N+1)/2$ ($N$ odd), and there are no phonons. }
             \label{Fig1}
\end{figure}

The Holstein Hamiltonian~\cite{Holstein59}, of a chain of $N$ sites (in the following $\hbar=1$), reads
\begin{equation}
H=-h\sum_{i=1}^{N-1}(c_{i}^{\dagger}c_{i+1}+h.c.)+g\sum_{i=1}^{N}(b_{i}+b_{i}^{\dagger})n_{i}+
\omega_{0}\sum_{i=1}^{N}b_{i}^{\dagger}b_{i}\label{HolsteinHam}.
\end{equation}
Here, $c_{i} (c_{i}^{\dagger})$ is the annihilation (creation) operator in the electron site $i$, and $b_{i}(b_{i}^{\dagger})$ is the phonon annihilation (creation) operator on the site $i$; $n_i=c^{\dagger}_ic_i$ is the electronic occupation number operator. The parameters $h$, $g$ and $\omega_0$ stand respectively for a nearest-neighbor (NN) site hopping for the electrons, electron-phonon coupling and free energy of the phonons. To encode the model in a trapped-ion chain, we first map the fermionic operators through the Jordan-Wigner transformation, $c_{i}\rightarrow\prod_{j=1}^{i-1}\sigma_{j}^{z}\sigma_{i}^{-}$ to tensor products of Pauli matrices. The mapped Hamiltonian describes now a coupled spin-boson system
\begin{eqnarray}
\nonumber H=h\sum_{i=1}^{N-1}(\sigma_{i}^{+}\sigma_{i+1}^{-}+h.c.)+\\ +g\sum_{i=1}^{N}(b_{i}+b_{i}^{\dagger})\frac{(\sigma_{i}^{z}+1)}{2}
+\omega_{0}\sum_{i=1}^{N}b_{i}^{\dagger}b_{i}.
\end{eqnarray}
The first term can be rewritten as $\frac{h}{2}\sum_{i=1}^{N}(\sigma_{i}^{x}\sigma_{i+1}^{x}+\sigma_{i}^{y}\sigma_{i+1}^{y})$. We now decompose the Hamiltonian into three parts, $H=H_1+H_2+H_3$. The single steps read
\begin{eqnarray}
 \label{steps}
\nonumber H_{1}=\sum^{N-1}_{i=1}\frac{h}{2}\sigma_{i}^{x}\sigma_{i+1}^{x}+\frac{\omega_{0}}{3}\sum_{i=1}^{N}b_{i}^{\dagger}b_{i},\\
H_{2}=\sum^{N-1}_{i=1}\frac{h}{2}\sigma_{i}^{y}\sigma_{i+1}^{y}+\frac{\omega_{0}}{3}\sum_{i=1}^{N}b_{i}^{\dagger}b_{i},\\ \nonumber H_{3}=\sum_{i=1}^{N}g(b_{i}+b_{i}^{\dagger})\frac{(\sigma_{i}^{z}+1)}{2}+\frac{\omega_{0}}{3}\sum_{i=1}^{N}b_{i}^{\dagger}b_{i}.
\end{eqnarray}
According to Ref.~\cite{Berry07}, one can upper bound the number of gates $N_g$ needed to achieve a simulation error smaller that $\epsilon$, by giving an upper bound for the norm of $H$~\cite{Suppl},
\begin{eqnarray}
N_{g} & \leq & 3\cdot5^{2k}[3(|h|(N-1)+2|g|N\sqrt{M-1}\nonumber\\&&+\omega_{0}NM)t]^{1+\frac{1}{2k}}/\epsilon^{1/2k}.
\end{eqnarray}
As mentioned before, the fractal depth $k$~\cite{Suzuki90} can be set to one in most applications. Here, we show the dependence of the number of gates in the number of fermionic sites $N$, and on the truncation in the number of bosons $M$. As the number of created phonons increases, one needs a higher-level truncation, and a larger Hamiltonian norm. Nevertheless, this shows that we can efficiently simulate a $2^{N}\times (M+1)^{N}$ Hilbert space, i.e., with a number of gates that grows at most polynomially in $N$ and $M$. To show the scaling of fidelities with the parameters considered, we plot in Fig.~\ref{Fig1} the time dependence of the fidelity loss $1-F(t)=1-|\langle\Psi_E(t)|\Psi_S(t)\rangle|^2$ of the simulated wavefunction $|\Psi_S(t)\rangle$ versus the exact one $|\Psi_E(t)\rangle$ as a function of coupling $g$ and of number of sites $N$. The particular decomposition has been chosen so that all terms in Eq. (\ref{steps}) can be implemented in a linear chain of trapped ions.

{\it Trapped-ion setup.- }  We consider a set of $N+1$ trapped ions in a chain, in order to simulate $N$ fermionic sites provided with Holstein interactions. The ions are bounded strongly in the radial direction, and confined longitudinally within a harmonic potential~\cite{James98}. We define $\nu_{i},\;i=1,2,...N+1$, as the frequencies of the axial normal modes. We relate the ion normal mode energies with the dispersionless phonon energies in Eq. (\ref{HolsteinHam}) via $\Delta_{i}=\nu_{i}-\frac{\omega_{0}}{3}$. The three Hamiltonian steps $H_{1}$, $H_{2}$ and $H_{3}$ are derived in the interaction picture with respect to
\begin{equation}
H_{0}=\sum_{i=1}^{N+1}\frac{\omega}{2}\sigma_{i}^{z}+\sum_{i=1}^{N}\Delta_{i}b_{i}^{\dagger}b_{i}+\nu_{N+1}b_{N+1}^{\dagger}b_{N+1}\label{eq:DefDeltai},
\end{equation}
where $\omega$ is the excitation energy of the individual ion taken as a two-level system, i.e., the carrier frequency.
In this way, the free energies of $N$ normal modes do not disappear in the interaction picture, and a flattered part of them is still present in order to recover the dispersionless phononic spectrum. 

\begin{figure}[t] 
\includegraphics[width=1\linewidth]{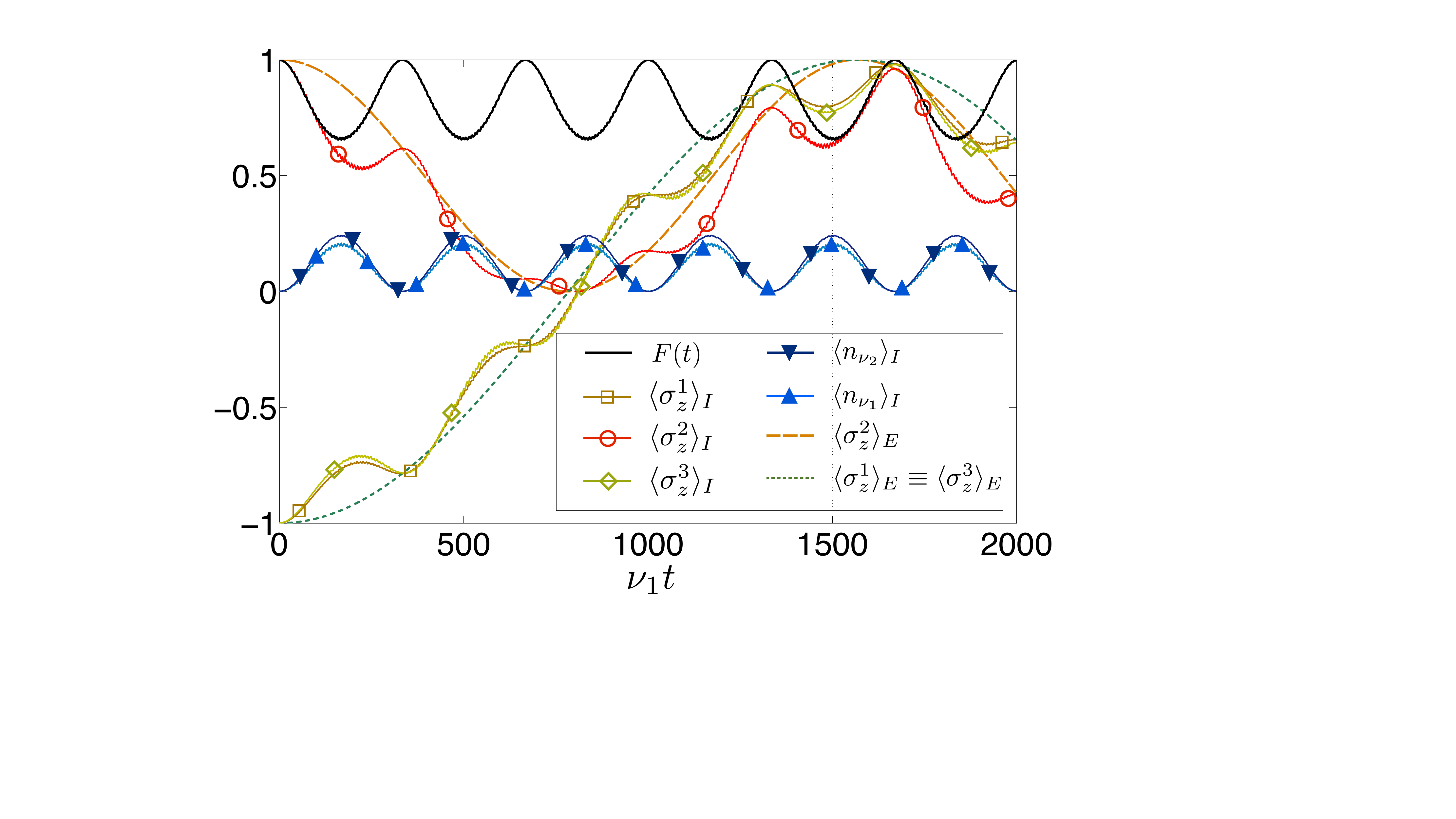}
\caption{(color online). Dynamics for the $3+1$ ions configuration of the NN XX Hamiltonian. Dotted curves stand for $\langle\sigma_z^{i}\rangle_{E}$ for the exact dynamics, and solid curves stand for $\langle\sigma_z^{i}\rangle_{I}$ for realistic ion interactions ($i=1,2,3$ for the first, second and third ion). The parameters are chosen in order to have maxima in the fidelity $F(t)=|\langle\Psi_{E}(t)|\Psi_{I}(t)\rangle|^2$ of $\sim 0.995$ (top black curve) at time steps of $\sim 333$~$\nu_1 t$. These time steps can be chosen as Trotter steps.}
             \label{Fig2}
\end{figure}

To simulate the dynamics associated to $H_1$ and $H_2$ of Eq. (\ref{steps}), one has to achieve a NN Ising coupling. The possibility of obtaining an Ising field in linear chains of trapped ions has been proposed and realized ~\cite{Kim09,Kim10}. However, in implementing NN interactions between more than two ions, one must be careful in designing an appropriate set of lasers and detunings in order to minimize the spurious non-nearest-neighbor (NNN) effects. To this extent, we have realized numerical simulations for a 3+1 ions setup~\cite{Suppl}, using one set of two pairs of counterpropagating lasers detuned close to the shifted center of mass (COM) shifted mode of frequency $\Delta_1=\nu_1-\omega_0/3$ to drive the first two ions (detunings $\pm \delta_1$), and another set of lasers detuned close to a second mode of frequency $\Delta_2$, that in the case of 3+1 ions can be chosen as the breathing mode, addressing the second and the third ion (detunings $\pm\delta_2$). For a generic number of ions, Rabi frequencies $\Omega_i$ of the lasers driving the $i$-th and the $i+1$-th ions are chosen to achieve the desired strength in the Ising coupling, according to~\cite{Kim09},
\begin{eqnarray}
\nonumber H_{NN}=\sum_{i=1}^{N-1}\Omega_{i}^2\left[ \left( \sum_{m=1}^{N}\frac{\eta_{i,m}\eta_{i+1,m}\Delta_{m}}{\delta_{i}^{2}-\Delta_{m}^{2}} \right)\right. \\
+\left. \frac{\eta_{i,N+1}\eta_{i+1,N+1}\nu_{N+1}}{\delta_{i}^{2}-\nu_{N+1}^{2}}\right] \sigma_{i}^{x}\sigma_{i+1}^{x}.
\end{eqnarray}
In Fig.  \ref{Fig2}, the first and second ion are driven with two pairs of counterpropagating lasers with detuning close to the shifted COM mode ($\delta_{1}=1.0187$~$\nu_{1}$ for $\omega_0=h/4$). The Rabi frequencies are chosen properly in order to reach a NN interaction of $h/2=0.001$~$\nu_1$.  Lasers driving the second and the third ions are detuned close to the shifted breathing mode at $\nu_{2}=1.731~\nu_1$~\cite{James98}, with parameters $\delta_{2}=1.71196$~$\nu_{1}$. Detunings are chosen  to have a dynamics decoupled with respect to the phonons at time steps $\sim333$~$\nu t$ and a negligible NNN interaction~\cite{Suppl}. At these times, the ion spins match the exact value, phonons are detached from spins and the fidelity oscillation (top black curve) $F(t)=|\langle\Psi_{E}(t)|\Psi_{I}(t)\rangle|^2$ reaches maxima, with peaks of $\sim0.995$. 

The initial state, as in all our numerical simulations, except where specified, is chosen to mimic a configuration in which one electron is injected at the center of a one dimensional lattice provided with Holstein interactions. To this extent, all the spins are initialized in the opposite Z direction, except the one in site $N/2$, in case of even $N$, or $(N+1)/2$ in case of odd $N$. The spin of the last ion has to be initialized along the Z direction in order to be a passive ion with respect to the dynamics, according to the protocol for the implementation of $H_3$ given below. The vibrational modes are assumed to be initially cooled down to the ground state with resolved sideband cooling~\cite{Leibfried03}.

Notice that one can always implement a perfect NN coupling by using more stroboscopic steps. A possibility is to decompose the global NN into nearest-neighbor pairwise interactions. Another possibility is to design a counter, non-nearest-neighbor interaction step between pairs of non-nearest neighbor ions in order to eliminate the spurious NNN imperfections. Given that one has an unwanted $h_{i,j}\sigma_x^i\sigma_x^{j}$, one can add more Trotter steps to the protocol of the form $-h_{i,j}\sigma_x^i\sigma_x^{j}$ in order to have an Hamiltonian free of NNN couplings. 
The dynamics associated to the step with $H_2$ is implemented similarly to the one of $H_1$, with a different choice of the initial phases of the lasers, in order to achieve a YY interaction.

The Hamiltonian $H_3$ is realized as a combination of $2N$ red and blue detuned lasers with appropriate initial phases in order to recover a coupling of the $i$-th ion ($i=1,...N$) with the $m_i$-th normal (shifted)  mode $\eta_{i,m_i}\Omega_i\sigma^i_x(b^{\dagger}_{m_i}+b_{m_i})$. The $i$-th ion is driven with red and blue detuned lasers to the $m_i$-th mode, establishing a one-to-one correspondence between the first $N$ ions and the first $N$ normal modes. Moreover, the last ion of the chain is driven by $2N$ lasers detuned in order to be coupled with the same modes of the ions in the chain. Two additional rotations of the spins of all ions around the Y axis are applied before and after coupling the spins to the phonons. They can be obtained by acting two times with a global beam upon all the $N+1$ ions at the same time. The Hamiltonian describing this process is,
\begin{equation}
H_{e-p}=\sum_{i=1}^{N}(\Omega_i\eta_{i,m_i}\sigma^i_z+\Omega_{N+1,i}\eta_{N+1,m_i}\sigma^{N+1}_z)(b_{m_i}+b^{\dagger}_{m_i}).
\end{equation}
The Rabi frequencies of the lasers must be chosen according to $\Omega_i=g/2\eta_{i,m_i}$, $\Omega_{N+1,i}=g/2\eta_{N+1,m_i}$.
If the last ion is initialized with the spin aligned along the Z axis and not addressed by spin flip gates during the simulation, the previous described gates result in the effective Hamiltonian on the {\it first $N$ ions subspace},
\begin{equation}
H_{e-p,N}=\sum_{i=1}^{N}g\frac{(\sigma^i_z+1)}{2}(b_{m_i}+b^{\dagger}_{m_i}).
\end{equation}

{\it Digital Simulation.-} In general, digital protocols are much sensitive to the state fidelity that one can achieve at the end of the digital step. According to the mathematical theory, increasing the number of steps will result in an increased fidelity on the final simulated state. However, if one has an error on a single step, increasing the number of gates will result in the accumulation of these errors. Thus on one hand the use of more accurate single gates is required, on the other hand one has to get a compromise between the increased fidelity due to the increased number of steps and the fidelity loss due to the accumulated single gate error.

\begin{figure}[t] 
\includegraphics[width=1\linewidth]{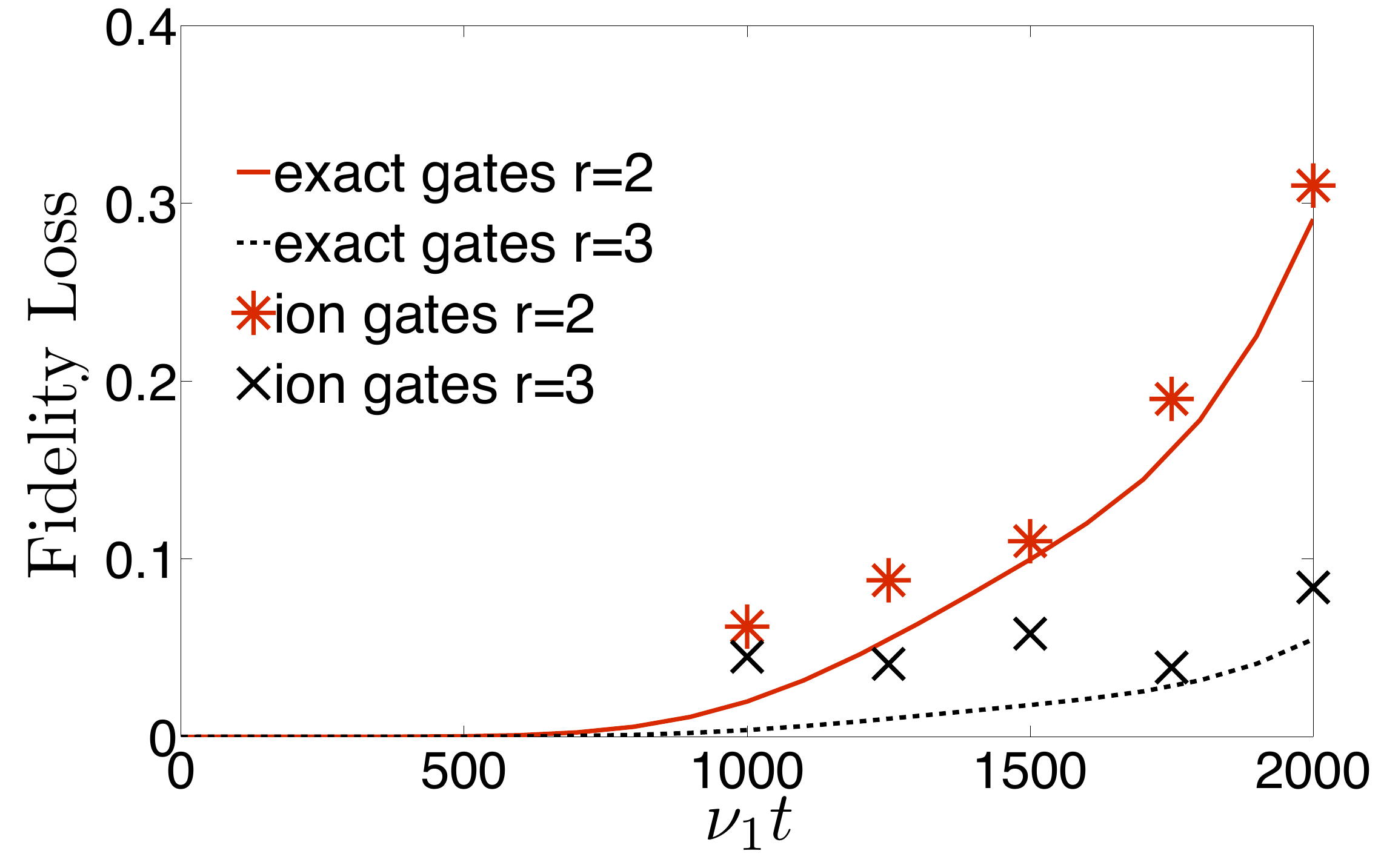}
\caption{(color online). Fidelity loss for 3+1 ion configuration, involving Trotter simulation with perfect gates and  realistic ion interactions, for two and three symmetric Trotter steps.}
             \label{Fig3}
\end{figure}

To have a quantitative estimation of the fidelity loss with the dynamics of the full ion Hamiltonian, we have realized numerical integrations for the Schr\"odinger equation for $N=$ 2+1~\cite{Suppl} and $N=$ 3+1 ion setups. We point out that we consider this reduced number of ions because of numerical computation restrictions, and to prove the feasibility of our model. In general our formalism may be straightforwardly extended to several ions. In Fig.\ref{Fig3}, a simulation for $r=2$ and $r=3$ symmetric Trotter steps is realized. The fidelity loss $1-|\langle\Psi_{E}(t)|\Psi_{S}(t)\rangle|^2$ for the Trotter protocol with perfect gates, i.e., associated to Hamiltonians $H_1$, $H_2$ and $H_3$, is plotted against points of fidelity loss $1-|\langle\Psi_{E}(t)|\Psi_{I}(t)\rangle|^2$ obtained with realistic trapped-ion gates including the full laser interactions  are plotted at various times. As can be appreciated, the fidelity loss for the ion gates is only slightly larger than for the exact Trotter gates, showing the feasibility of the protocol with realistic trapped-ion interactions. The total simulation time has been chosen in order to remain under the decoherence time for the ions~\cite{Haffner08}. The frequency of the center of mass mode can be assumed to be $\nu_1\simeq 2\pi\times1$~MHz. The global rotation for the ion spins can be assumed to be done in $7$~$\mu$s~\cite{Lanyon11}. The number of global rotations is $4r$. The step for the red and blue sideband Hamiltonian can be performed in the same time as the step for the NN XX gate (or even faster). Provided with these parameters, for a final simulated time of $2000/\nu_1\sim318~\mu $s, the time spent for the simulation can be taken of $\sim1$~ms. Given that typical heating rates in trapped ion experiments~\cite{Lanyon11} are of about 1 phonon/s, we can assume that for the time of the proposed simulation heating will not be significant. 

Tuning the coupling strength $g$ by setting the Rabi frequencies of the red and blue detuned lasers to various values, one can measure the different correlations between electron and phonon displacement at distant sites,
\begin{equation}
\chi (i,j)=\langle\Psi(t)|c^{\dagger}_ic_i(b_j^{\dagger}+b_j)|\Psi(t)\rangle.
\end{equation}
This will amounts to a signature of the polaron size~\cite{Romero99}. Ranging from small to large $g$ will lead to a measure of the crossover between large/small polaron.  Notice that these correlations are mapped in our ion setup onto
\begin{equation}
\chi (i,j)=\langle\Psi(t)|(b_{m_j}+b_{m_j}^{\dagger})\frac{(\sigma_{i}^{z}+1)}{2}|\Psi(t)\rangle,
\end{equation}
which can be measured by mapping the motional onto the internal state of the auxiliary $N+1$-th ion, and then detecting resonance fluorescence of ions $N+1$ and $i$~\cite{Gerritsma1,Gerritsma2}.
We notice that with our setup the possibility of simulating a 2D and 3D Holstein model is provided, by encoding two and three dimensional interactions into a linear chain by addressing distant ions with nonlocal gates~\cite{Casanova12}.

 Currently, more than 100 gates have been implemented in a trapped-ion quantum simulation experiment with Trotter methods~\cite{Lanyon11}. In the near future, it should be possible to achieve hundreds or even thousands of gates per experiment~\cite{RainerPrivate}, allowing our proposal to reach about ten qubits. It is noteworthy to mention that our proposed digital quantum simulation will already overcome the limits of classical computers with 10 ions and 5 phonons per ion. This will allow to study the formation of small polarons under these conditions. Future experiments involving 20 to 30 ions will permit to address the study of more complex dynamics, including electron-electron correlations mediated by phonons. In this manner, the trapped-ion quantum simulator will prove to be a remarkable tool for simulating fermions coupled to bosons and related condensed-matter or high-energy physics scenarios.
 
The authors acknowledge funding from Basque Government BFI08.211 and IT472-10, EC Marie Curie IEF grant, Spanish MICINN FIS2009-12773-C02-01, UPV/EHU UFI 11/55, SOLID, CCQED, and PROMISCE European projects.

\pagebreak

\section{Supplemental Material for "Digital Quantum Simulation of the Holstein Model in Trapped ions"}
\maketitle

\section{Upper bound for the norm}

In this section, we give an upper bound for the the norm of $H$ in Eq.~(3) of the main text, in order to bound the error one makes with a Suzuki-Lie-Trotter expansion~\cite{Suzuki90S}. Consequently, we bound the number of gates one needs for achieving a given fidelity on the simulated quantum state.
The norm is bounded by the sum of the norms of each term appearing in $H$. The computation of single norms amounts to finding the largest eigenvalue of the single terms
\begin{eqnarray}
||H||\leq|h|\cdot\sum_{i=1}^{N-1}\left\Vert (\sigma_{i}^{+}\sigma_{i+1}^{-}+h.c.)\right\Vert +\\
\nonumber +|g|\sum_{i=1}^{N}\left\Vert (b_{i}+b_{i}^{\dagger})\frac{(\sigma_{i}^{z}+1)}{2}\right\Vert +\omega_{0}\sum_{i=1}^{N}\left\Vert b_{i}^{\dagger}b_{i}\right\Vert .
\end{eqnarray}
Let us consider the various norms separately. The term expressed by $|h|\sum_{i=1}^{N-1}\left\Vert (-\sigma_{i}^{+}\sigma_{i+1}^{-}+h.c.)\right\Vert=|h/2|\sum_{i=1}^{N-1}\left\Vert (\sigma_{i}^{x}\sigma_{i+1}^{x}+\sigma_{i}^{y}\sigma_{i+1}^{y})\right\Vert  $ represents a sum of $2(N-1)$ tensor products of Pauli matrices, with norm $1$. Therefore, $|h|\sum_{i=1}^{N-1}\left\Vert (-\sigma_{i}^{+}\sigma_{i+1}^{-}+h.c.)\right\Vert \leq |h|(N-1)$. The norm $\left\Vert b_{i}^{\dagger}b_{i}\right\Vert $ is bounded
by the truncation in the number of bosons in the mode $i$ , as is clear by the standard Fock representation
\begin{equation}
b_{i}^{\dagger}b_{i}\rightarrow\left(\begin{array}{cccc}
0\\
 & 1\\
 &  & \ddots\\
 &  &  & M
\end{array}\right),\;\left\Vert b_{i}^{\dagger}b_{i}\right\Vert =M.
\end{equation}
Therefore $\sum_{i=1}^{N}\left\Vert b_{i}^{\dagger}b_{i}\right\Vert =\sum_{i=1}^{N}M=NM$.

The norm $\sum_{i=1}^{N}\left\Vert (b_{i}+b_{i}^{\dagger})\frac{(\sigma_{i}^{z}+1)}{2}\right\Vert $
is equivalent to $\sum_{i=1}^{N}\left\Vert (b_{i}+b_{i}^{\dagger})\right\Vert $ , given that $\left\Vert \frac{(\sigma_{i}^{z}+1)}{2}\right\Vert =1$.
The term $(b_{i}+b_{i}^{\dagger})$ in the Fock basis reads 
\begin{equation}
(b_{i}+b_{i}^{\dagger})\rightarrow\left(\begin{array}{cccccc}
0 & 1\\
1 & 0 & \sqrt{2}\\
 & \sqrt{2} & 0\\
 &  &  & \ddots\\
 &  &  &  & 0 & \sqrt{M}\\
 &  &  &  & \sqrt{M} & 0
\end{array}\right).
\end{equation}
The characteristic polynomial of the matrix for the truncation to $M$ bosons is given
in a recursively way,
\begin{eqnarray}
\nonumber D_0(\lambda)=1,\; D_1(\lambda)=-\lambda \\
D_n(\lambda)=-\lambda D_{n-1}(\lambda)-(n-1)D_{n-2}(\lambda).
\end{eqnarray}
The $D_n(\lambda)$ are a $\sqrt{2}$ rescaled version of the Hermite polynomials. A simple bound for the largest zero of $D_M(\lambda)$ (i.e. the norm of the bosonic displacement operator) is given by the expression $2\sqrt{M-1}$ (see for example Ref.~\cite{Area04S}). Summarizing, the norm for the Holstein Hamiltonian is upper bounded by
\begin{equation}
\left\Vert H\right\Vert \leq|h|(N-1)+2|g|N\sqrt{M-1}+\omega_{0}NM.
\end{equation}

\section{Numerical simulations}

In this section, we provide additional numeric plots and further discussions for our simulation protocol for a 2+1 and 3+1 ion configuration. First of all, in all our numerical simulations, we have fixed the total {\it simulated} time to a maximum of $2000/\nu_1$, in units of the center of mass (COM) mode frequency $\nu_1$. Assuming $\nu_ 1\sim2\pi\times1$~MHz, this gives a total simulated time of $\sim318$~$\mu$s. We choose Trotter steps equally extended within a time $\tau=t/2$, where $t$ is the total simulated time. With these assumptions, to compute the total effective {\it simulation} time, one has to multiply the simulated time by the   
number of terms in the decomposition of the simulated dynamics, i.e. 3 in our case. Then we have to add the time contribution for the global $\pi /4$ rotations along the Y axis, necessary to achieve the Z-like coupling to the phonons, that can be estimated to be around $\sim7$~$\mu$s each~\cite{Lanyon11S}. 
Considering four global rotations per symmetric Trotter step, this gives a total simulation time of the order $\sim1$~ms for the $r=1$ and $r=2$ case. This is well below the typical decoherence times for a trapped-ion setup~\cite{Haffner08S}. Notice that we have made assumptions on the time extension for the Trotter steps, but nothing prevents to set the duration for the Trotter step to shorter times, as long as one can adjust properly the Rabi frequencies of the lasers used~\cite{Lanyon11S}. This paves the way to the scalability of the protocol.

\begin{figure}
\includegraphics[width=1\linewidth]{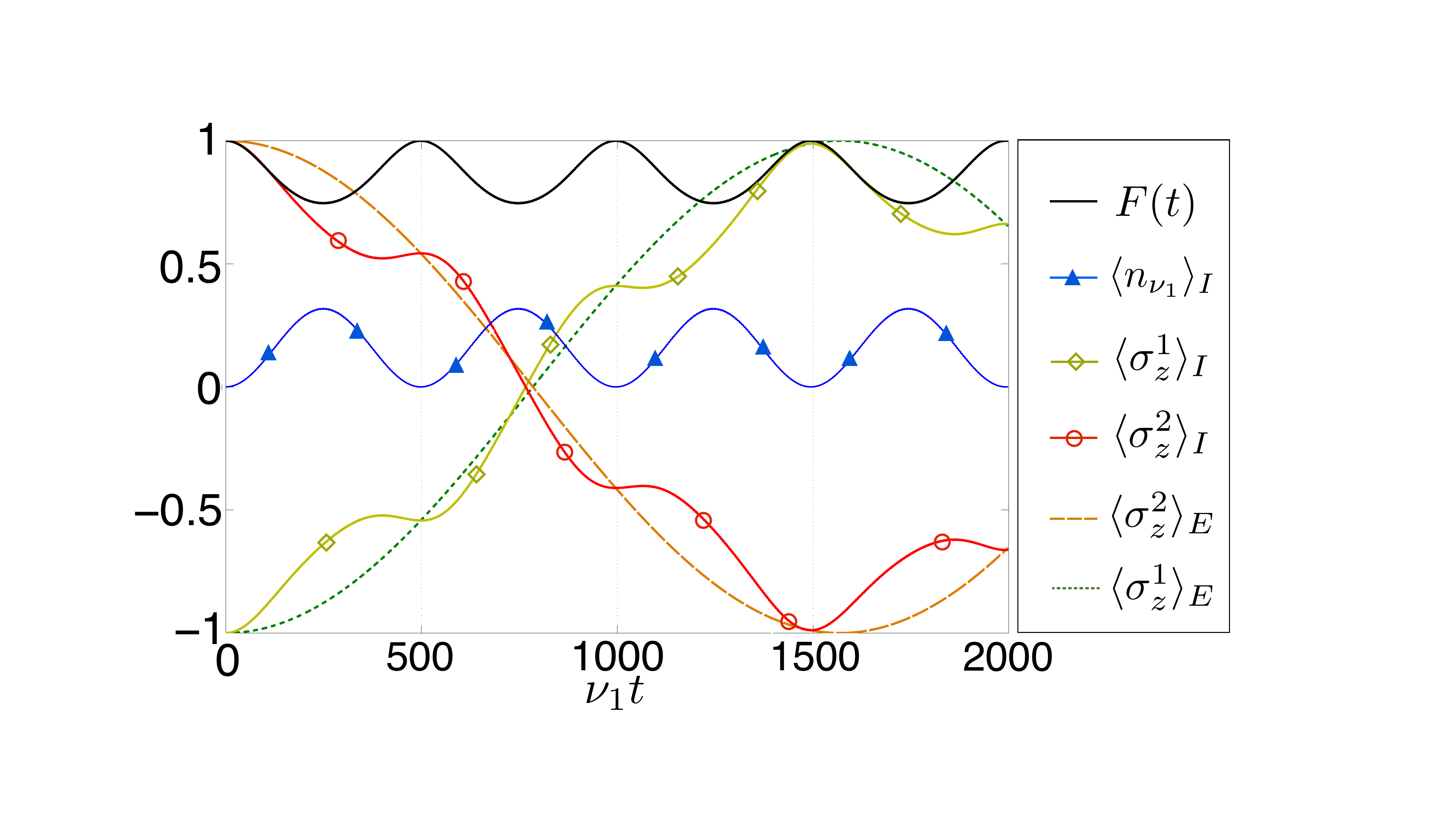}
\caption{(color online). Dynamics for the $2+1$ ions configuration of the NN XX Ising Hamiltonian. Dotted curves stand for $\langle\sigma_z^{i}\rangle_{E}$ for the exact dynamics, and solid curves stand for $\langle\sigma_z^{i}\rangle_{I}$ for realistic ion interactions ($i=1,2$ for the first and second ion). The parameters are chosen in order to have fidelity losses of $1-F(t)=1-|\langle\Psi_{E}(t)|\Psi_{I}(t)\rangle|^2\sim10^{-4}$ (top black curve) at time steps of $\sim 500/\nu_1$.}
             \label{Fig4}
\end{figure}

The dynamics described by Hamiltonians $H_1$ and $H_2$ can be achieved by using two pairs of counterpropagating lasers with opposite detunings $\pm\delta_i$~\cite{Kim09S}, driving the $i$-th and the $i+1$-th ions. One can switch between a nearest neighbor (NN) XX/YY Ising interactions by taking appropriate initial phases for the lasers. The effective spin-spin coupling generated by this kind of laser drivings has the form of Eq.~(7) in the main text.

In order to have negligible phonon displacements at the Trotter step time $\tau$, one has to choose the detuning $\delta_i=\pm2 \pi/\tau +\Delta_m$ close to one of the modes of (shifted) frequencies $\Delta_m$ (thus $|\delta_i-\Delta_m|\ll|\Delta_m|,|\delta_i|$). We point out again here that in our protocol we deal with shifted frequencies, as explained in the main text, to take into account the desired dispersionless energies of the Holstein phonons. The  $\pm$ sign in the choice of $\delta_i$ can be used to change the relative sign of the spin-spin interaction, depending on $\sgn[\eta_{i,m}\eta_{i+1,m}]$. We assume in our simulations a relative Lamb-Dicke parameter distribution for ions and modes as in~\cite{James98S}, with an overall magnitude of $0.1$. If one chooses $m=1$ for the 2+1 ions setup, i.e. a detuning close to the COM mode, it must be set to $\delta=2\pi/\tau+\Delta_1$, to obtain a positive $h/2$ coupling in the Ising NN interaction, because $\sgn[\eta_{1,1}\eta_{2,1}]=+$ .  

In Fig.~\ref{Fig4}, we show the numerical integration for the dynamics of the NN XX Ising interaction for a 2+1 ions configuration. The simulated strength for the Ising coupling is $h/2=0.001$~$\nu_1$. For $\omega_0=h/4$, one has a shifted frequency for the COM mode of $\Delta_1=\nu_1-0.0005\nu_1/3$. By choosing $\tau=500/\nu_1$, the detuning used in this case is $\delta=2 \pi\nu_1/500+\Delta_1=1.0124$~$\nu_1$, i.e. for $\nu_1\simeq 2\pi \times 1$~MHz, a frequency difference with the mode of   $\delta-\nu_1=2\pi\times12.4$~KHz. The Rabi frequencies of the lasers are chosen in order to recover the desired strength for the Ising coupling.

\begin{figure}[t] 
\includegraphics[width=1\linewidth]{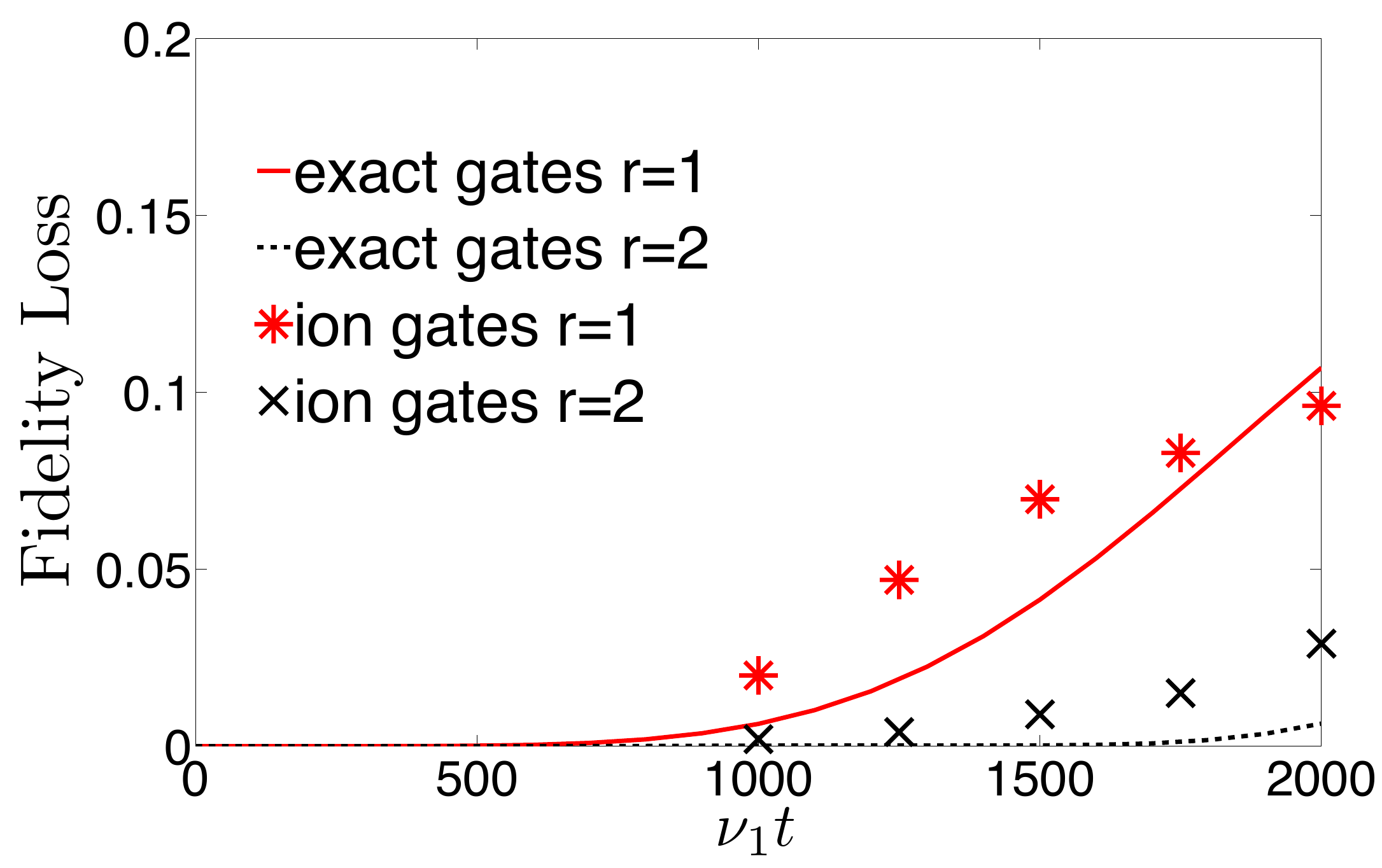}
\caption{(color online). Fidelity loss for a 2+1 ions configuration as a function of time and Trotter steps. The simulated Holstein interaction has parameters $g=h/10$, $\omega_0=h/4$. Dotted and solid lines stand respectively for a simulation with one and two symmetric Trotter steps. Single points stand for the error in the simulation protocol using ion gates. }
             \label{Fig5}
\end{figure}

To have an idea of how real ion interactions affect the protocol for a 2+1 ion setup, we make a plot of the errors on the simulated state with perfect gates and with ion gates in Fig.~\ref{Fig5}. One clearly sees that the higher fidelities obtained by using the ion gates with respect to the 3+1 ion setup are due to the higher single gate fidelity for the 2+1 setup, which permits to explore better fidelity regimes. The simulated parameters here are $g=h/10$, $\omega_0=h/4$, $h=0.002$~$\nu_1$. We remark that in the simulations we have used a small $g/h$ ratio to reduce the complexity of the simulation (i.e., the necessary truncation for the Fock space is small). Nevertheless, in a trapped-ion experiment, big $g/h$ ratios with large freedom for the choice of $\omega_0$ can be explored, thus recovering the typical {\it self trapping line} for the formation of small polarons~\cite{Romero99S}. The time points for the simulation range from $t=1000/\nu_1$ to $t=2000/\nu_1$. For $r=1$ this gives Trotter steps ranging from $\tau=500/\nu_1$ to $\tau=1000/\nu_1$. The detuning for the NN interaction has to be set accordingly at each point, ranging from $1.0124$~$\nu_1$ to $1.0061$~$\nu_1$. 

\begin{figure}[t] 

\includegraphics[width=1\linewidth]{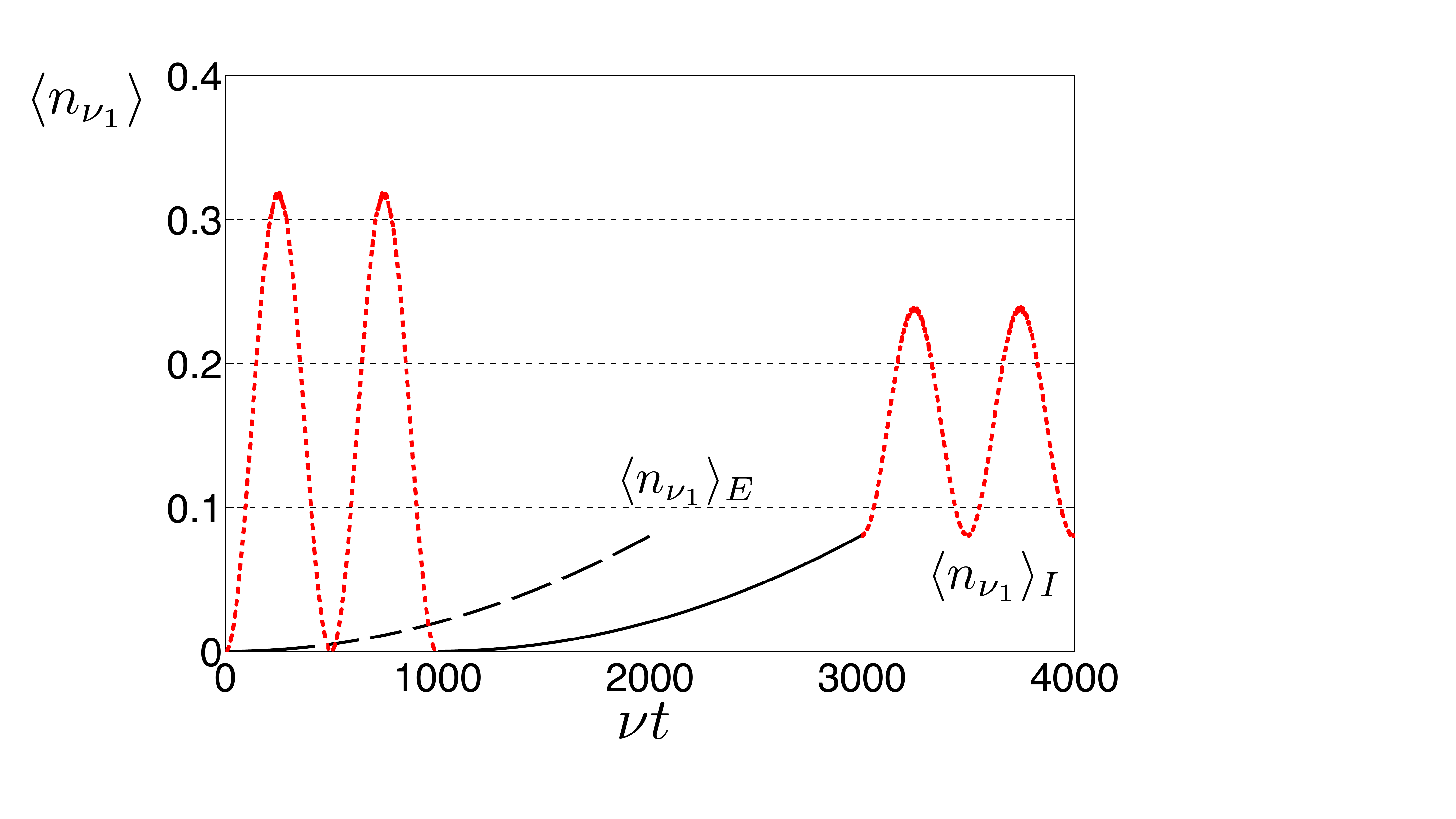}
\caption{(color online). Mean number of phonons inside a Trotter protocol for a 2+1 ions configuration, for the Hamiltonian $H=H_1+H_3$, where $H_1$ is a spin spin XX interaction and $H_3$ is a spin-phonon coupling interaction, as in Eq.(4) in the main text. The phonons are excited within the time for the $H_3$ steps (solid black line), and excited and released to their initial value within the $H_1$ interactions (dotted red lines), with the typical oscillations for this kind of gates. }
             \label{Fig6}
\end{figure}

To obtain the plot of Fig.~3 in the main paper, with the same simulated parameter $g=h/10$, $\omega_0=h/4$, $h=0.002$~$\nu_1$, we have used two simultaneous NN XX interactions as described above. One involves the first two ions with a detuning close to the COM mode, and another one driving the second and the third ion with a detuning close to the breathing mode, of frequency $\nu_2=1.731$~$\nu_1$~\cite{James98S}. To obtain an Ising interaction with the proper sign for the second and the third ion, we have to set the detuning for the second laser below the shifted frequency $\Delta_2$. This is because for a 3+1 configuration (i.e., four ions in a linear trap), one has $\sgn [\eta_{1,1}\eta_{2,1}]\neq \sgn [\eta_{2,2}\eta_{3,2}]$~\cite{James98S}.
The detuning for the second set of lasers is therefore chosen to be $\delta=-2\pi/\tau+\Delta_2$. For example, for the simulation point at $t = 1000/\nu_1$, corresponding to $\tau=250/\nu_1$ for $r=2$, one has $\delta_1=1.025$~$\nu_1$ and $\delta_2=1.7057$~$\nu_1$.
Using these parameters, it turns out that the non-nearest-neighbor coupling between the first and the third ion is negligible.

To get an insight of what happens to the phonon population of the COM mode inside a Trotter protocol, one can have a look to Fig.~\ref{Fig6}. Here, it is shown the mean number of phonons for a 2+1 ion setup using a symmetric decomposition at $r=1$ of the Hamiltonian $H=H_1+H_3$, where $H_1$ is a XX Ising interaction obtained with a detuning close to $\Delta_1$ and $H_3$ is a Z-like coupling to phonons. The decomposed evolution operator has the form $U_2(t)=e^{-iH_1t/2}e^{-iH_3t/2}e^{-iH_3t/2}e^{-iH_1t/2}$.
 We see that, in the first and the last step, the two Ising interactions create phonons, while relaxing them at the end of the step, because the laser detuning and Rabi frequencies are chosen to obtain detachment from the phonons at the end of the Trotter step. In the two middle steps, the phonons are excited according to the $H_3$ Hamiltonian. The final mean value for the phonon number is recovered with respect to the dashed line value, which is the numerical value according to the exact evolution operator $e^{-iHt}$, with an error of $\sim1\%$. Notice that since the decomposition involves symmetric Trotter steps, and each one is chosen to be of the same duration, the total simulation time is doubled.

\section{Estimation of Non-nearest-neighbor couplings}

In this paragraph we give an estimation of the NNN couplings that appear when one wants to generate the NN Ising interaction with the parameters that we use in the main text. We propose to address pairs of NN ions with independent counterpropagating couples of lasers detuned close to different modes, i.e. a different mode is assigned to a specifiic couple of NN ions. This gives rise to NNN coupling between distant ions, which we show being negligible for specific detunings and gate times. For a 3+1 ion configuration, for example, the total Hamiltonian is 
\begin{widetext}
\begin{equation}
\label{crossHam}
H=H_{1}+H_{2}=\sum_{m}\sin(\delta_{1}t)\left(a_{m}e^{-i\nu_{m}t}+a_{m}^{\dagger}e^{i\nu_{m}t}\right)\sum_{i=1}^{2}\Omega_{i}\eta_{i,m}\sigma_{i}^{x}+\sum_{m}\sin(\delta_{2}t)\left(a_{m}e^{-i\nu_{m}t}+a_{m}^{\dagger}e^{i\nu_{m}t}\right)\sum_{i=2}^{3}\Omega_{i}\eta_{i,m}\sigma_{i}^{x},
\end{equation}
\end{widetext}
obtained by driving the first two ions with two pairs of counterpropagating lasers detuned to $\pm\delta_1$~\cite{Kim09S}, while the lasers driving the second and the third ion are detuned to $\pm\delta_2$. Therefore a second order Magnus expansion of the Hamiltonian in Eq.(\ref{crossHam}) leads to unwanted NNN terms in the evolution operator of the form
\begin{widetext}
\begin{equation}
\label{Magnus}
\left(\int_{0}^{t}dt'\int_{0}^{t'}dt''[H_{1}(t'),H_{2}(t'')]\;+\;\int_{0}^{t}dt'\int_{0}^{t'}dt''[H_{2}(t'),H_{1}(t'')]\right)=\sum_m\left(Z_{1,m}(t)+Z_{2,m}(t)\right)S_{1,m}S_{2,m},
\end{equation}
\end{widetext}
where we have defined $S_{1,m}=\sum_{i=1}^{2}\Omega_{i}\eta_{i,m}\sigma_{i}^{x}$, $S_{2,m}=\sum_{i=2}^{3}\Omega_{i}\eta_{i,m}\sigma_{i}^{x}$. Some straightforward algebra leads to 
\begin{widetext}
\begin{equation}
\label{Zeta1}
Z_{1,m}(t)=\frac{i}{2(\delta_{1}^{2}-\nu_{m}^{2})}\left(\delta_{1}\frac{\sin(\delta_{2}-\nu_{m})t}{(\delta_{2}-\nu_{m})}-\delta_{1}\frac{\sin(\delta_{2}+\nu_{m})t}{(\delta_{2}+\nu_{m})}+\nu_{m}\frac{\sin(\delta_{2}-\delta_{1})t}{(\delta_{2}-\delta_{1})}-\nu_{m}\frac{\sin(\delta_{2}+\delta_{1})t}{(\delta_{2}+\delta_{1})}\right),
\end{equation}
\begin{equation}
\label{Zeta2}
Z_{2,m}(t)=\frac{i}{2(\delta_{2}^{2}-\nu_{m}^{2})}\left(\delta_{2}\frac{\sin(\delta_{1}-\nu_{m})t}{(\delta_{1}-\nu_{m})}-\delta_{2}\frac{\sin(\delta_{1}+\nu_{m})t}{(\delta_{1}+\nu_{m})}+\nu_{m}\frac{\sin(\delta_{1}-\delta_{2})t}{(\delta_{1}-\delta_{2})}-\nu_{m}\frac{\sin(\delta_{1}+\delta_{2})t}{(\delta_{1}+\delta_{2})}\right).
\end{equation}
\end{widetext}
These contributions are negligible for the parameters that we use, i.e. first detuning close to the first mode and second detuning close to the second one, $|\nu_1-\delta_1|\ll\nu_1$, $|\nu_2-\delta_2|\ll\nu_2$. For example, taking the strongest resonant
term from the series in Eq. (\ref{Magnus}), $m=2$ for $Z_{2,2}(t)$, the first term on the right side in Eq. (\ref{Zeta2}) reads
\begin{equation}
\frac{\delta_{2}\sin(\delta_{1}-\nu_{2})t}{2(\delta_{2}^{2}-\nu_{2}^{2})(\delta_{1}-\nu_{2})}=\frac{\delta_{2}\sin(\delta_{1}-\nu_{2})t}{2(\delta_{2}+\nu_{2})(\delta_{2}-\nu_{2})(\delta_{1}-\nu_{2})}.
\end{equation}
Since $\Omega_i\eta_{i,m}\cong\Omega_j\eta_{j,n}$, the term is negligible in comparison to the desired NN terms, whose
couplings goes like $-\frac{i\nu_{2}t}{2(\delta_{2}^{2}-\nu_{2}^{2})}$, $-\frac{i\nu_{1}t}{2(\delta_{1}^{2}-\nu_{1}^{2})}$,
for sufficient large times, 
\begin{equation}
\Biggl\lvert \frac{\nu_{2}t}{(\delta_{2}+\nu_{2})(\delta_{2}-\nu_{2})}\Biggl\rvert \gg \Biggl\lvert  \frac{\delta_{2}}{(\delta_{2}-\nu_{2})(\delta_{2}+\nu_{2})(\delta_{1}-\nu_{2})}\Biggl\rvert,
\end{equation}
\begin{equation}
\label{timecond}
t\gg \Bigg \lvert \frac{\delta_{2}}{\nu_{2}}\frac{1}{(\delta_{1}-\nu_{2})}\Bigg \rvert .
\end{equation}
For realistic parameters the critical time is $t\sim~1~/\nu_1$. Since our gates are
obtained at times $\tau\sim~100/\nu_1$, these NNN terms can be neglected. We stress again that in the protocol some of the frequencies $\nu_i$ have to be shifted, we have left the original frequencies to avoid a heavy notation. Same kind of considerations are valid for the other terms in the right side of Eq. (\ref{Zeta2}) and Eq. (\ref{Zeta1}). This also extends in a straightforward way to couplings between any two NNN ions in a configuration with an arbitrary number of ions, as long as conditions like Eq. (\ref{timecond}) are satisfied.

\end{document}